\documentclass{mem}
\usepackage{natbib}\usepackage{txfonts}\usepackage{balance}
\usepackage{graphicx}
\usepackage[a4paper,breaklinks,dvipdfm]{hyperref}
\idline{75}{282}
\begin{document}
\def\teff{$T\rm_{eff }$}
\def\kms{$\mathrm {km s}^{-1}$}

\title{
Role of ejecta clumping and back-reaction of accelerated cosmic rays in
the evolution of supernova remnants
}

   \subtitle{}

\author{
S. \,Orlando\inst{1} \and F. \,Bocchino\inst{1} \and M.
\,Miceli\inst{2,1} \and O.  \,Petruk\inst{3} \and M. L. \,Pumo\inst{4}
          }

  \offprints{S. Orlando}
 
\institute{
  INAF -- Oss. Astronomico di Palermo, Piazza del Parlamento 1,
  90134 Palermo, Italy
\and
  Dip. di Fisica, Univ. di Palermo, Piazza del Parlamento 1, 90134 Palermo,
  Italy
\and
  Inst. for Applied Problems in Mechanics and
  Mathematics, Naukova St. 3-b Lviv 79060, Ukraine
\and
  INAF -- Oss. Astronomico di Padova, Vicolo dell'Osservatorio 5,
  35122 Padova, Italy\\
\email{orlando@astropa.inaf.it} }

\authorrunning{Orlando et al.}

\titlerunning{Ejecta clumping and back-reaction of accelerated cosmic
rays in SNRs}

\abstract{
The thermal structure of the post-shock region of a young supernova
remnant (SNR) is heavily affected by two main physical effects, the
back-reaction of accelerated cosmic rays (CRs) and the Rayleigh-Taylor
(RT) instabilities developing at the contact discontinuity between the
ejecta and the shocked interstellar medium (ISM). Here, we investigate
the role played by both physical mechanisms in the evolution of SNRs
through detailed 3D MHD modeling. Our model describes the expansion of
the remnant through a magnetized ISM, including consistently the initial
ejecta clumping and the effects on shock dynamics due to back-reaction
of accelerated CRs. We discuss the role of the initial ejecta clumpiness
in developing strong instabilities at the contact discontinuity which
may extend upstream to the main shock and beyond.
}
\maketitle{}

\section{Introduction}

SNRs are the site where CR diffusive shock acceleration
occurs. Indirect evidence of CRs acceleration comes from the separation
between the forward shock and the contact discontinuity. In SN\,1006, 
observations have shown that the azimuthal profile of the ratio between
the forward shock and the contact discontinuity radii $R_{\rm bw}/R_{\rm
cd}$ is fairly uniform and much lower than predicted for a non-modified
shock (\citealt{Miceli09}). Recently \citet{Rakowski11} have found and
analyzed clumps of ejecta close to or protruding beyond the main blast
wave. These findings have been interpreted as a consequence of the energy
losses to CRs at the forward shock. However, several authors have shown
that extreme energy losses to accelerate the CRs are needed to allow
a significant fraction of the ejecta to approach or even overtake the
forward shock (e.g \citealt{Miceli09}). In addition, the observations
show that $R_{\rm bw}/R_{\rm cd}$ is lower than predicted by non-modified
shock models even in regions dominated by thermal emission where the CR
acceleration efficiency is supposed to be low (e.g. \citealt{Miceli09}).

An alternative in the data interpretation could be the ejecta structure
of the explosion itself.  In fact, density inhomogeneities in the ejecta
can enhance the growth of RT instabilities potentially
allowing them to reach the forward shock. The question is: can the ejecta
clumping enhance the growth of RT instabilities up to a level that allows
them to reach and possibly overtake the forward shock? In fact, there
is a growing consensus that density clumping of ejecta is intrinsic at
early phases of the remnant evolution. The density structure of ejecta
therefore may naturally explain the small values of $R_{\rm bw}/R_{\rm
cd}$ observed in SNRs.

Here we investigate this issue by developing a 3D MHD model describing
the expansion of a SNR through a magnetized medium, including 
consistently both the initial ejecta clumping and the effects on shock
dynamics due to back-reaction of accelerated CRs.

\section{Three-dimensional MHD Modeling}

We improved the 3D MHD model discussed by \citet{Orlando07,
Orlando11}. The SNR is modeled by numerically solving the time-dependent
MHD equations of mass, momentum, and energy conservation, including
consistently the initial ejecta clumping and the effects on shock
dynamics due to back-reaction of accelerated CRs (Orlando et al. 2011,
in preparation). The calculations were performed using FLASH, an
adaptive mesh refinement multiphysics code for astrophysical plasmas
\citep{Fryxell00} extended with additional computational modules to
handle the back-reaction of accelerated CRs. The latter process is
included in the model by following the approach of \citet{Ferrand10} and
extending their method to MHD models. In particular, our model includes
an effective adiabatic index $\gamma_{\rm eff}$ evolving in time: at each
time-step of integration the time- and space-dependent $\gamma_{\rm eff}$
is calculated at the shock front and then is advected within the remnant,
remaining constant in each fluid element. In addition to the approach of
\citet{Ferrand10} we allow $\gamma_{\rm eff}$ to depend on local physical
conditions and on the injection rate of particles, by assuming that the
index depends on the obliquity angle between the upstream magnetic field
and the normal to the shock.

As for the density structure of the ejecta, we have investigated
two different initial ejecta density profiles: the exponential
profile that has been shown to be the most representative of explosion
models for thermonuclear SNe (\citealt{Dwarkadas98}), and the power-law
profile with $n=7$ that has been used to represent deflagration models
(\citealt{Nomoto84}). We assume that the initial ejecta have a clumpy
structure, modeled by defining small clumps including a perturbation of
the mass density. For each of the initial profiles, we explore density
clumps of ejecta with initial size either 2\% or 4\% of the initial
remnant radius. Initially each clump has a random density contrast
derived from an exponential distribution that is characterized by a
maximum density contrast. We explore maximum contrasts ranging
between 1.5 and 5.

As for the reference physical system, we adopted parameters appropriate
to describe SN\,1006: density of ISM $n_{\rm ism} = 0.05$ cm$^{-3}$,
energy of the SN explosion $E_{\rm SN} = 1.5\times 10^{51}$ ergs, mass
of ejecta $M_{\rm ej} = 1.4~M_{\odot}$, shock velocity after 1000 yrs
$V_{\rm shock} = 5000$ km s$^{-1}$, and diameter of the remnant after
1000 yrs $D = 17$ pc. The initial magnetic field configuration is that
suggested by \citet{Bocchino11} in the case of SN\,1006 and resulting
from the comparison of radio observations with MHD models. The initial
radius of the remnant is 0.5 pc. We follow the remnant evolution for
1000 yrs. Special emphasis was placed on capturing the enormous range in
spatial scales in the remnant: in fact we need to perform simulations with
sufficient spatial resolution to capture the structures of the ejecta in
the initial system configuration and to follow the evolutions of these
structures. To this end, we exploited the adaptive mesh capabilities of
the FLASH code by using 11 nested levels of resolution; the effective
mesh size was $8192\times8192\times8192$.

\section{Results}

\begin{figure}[t!]
\resizebox{\hsize}{!}{\includegraphics[clip=true]{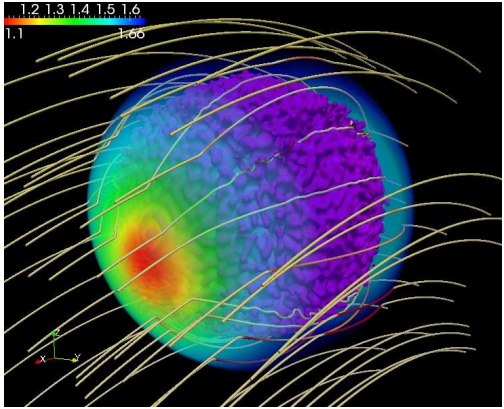}}
\caption{\footnotesize
3D volume rendering describing the spatial distribution of the effective
adiabatic index for a model with an exponential profile of the initial
ejecta density after 1000 yrs. The index $\gamma_{\rm eff}$ is minimum
in red regions (see color bar). The violet surface tracks the ejecta
material and the white lines are sampled magnetic field lines.
}
\label{fig1}
\end{figure}

We first analyzed the back-reaction of accelerated CRs, by
considering models accounting for the shock modification but without
initial clumping of ejecta. As an example, Fig.~\ref{fig1} presents the
results for a model with an exponential profile of the initial ejecta
density after 1000 yrs of evolution. In this model we have assumed that
the back-reaction of accelerated CRs is modulated by the obliquity angle,
being more efficient at parallel shocks where the effective adiabatic
index is minimum, namely $\gamma_{\rm eff} = 1.1$ (we are assuming extreme
energy losses to accelerate the CRs).

We investigated the effect of accelerated CRs on the separation
between the blast wave and the contact discontinuity. To this end,
we derived the azimuthal profiles of the ratio between the blast wave
and contact discontinuity radii $R_{\rm bw}/R_{\rm cd}$ from the models
and compared them with that derived from the observations of SN\,1006
(\citealt{Miceli09}). In general, we found that the modeled profiles are
higher than those observed and are modulated by the obliquity angle (see
black line in Fig.~\ref{fig4}).  The observations can be reproduced only
in limited regions where the effects of accelerated CRs is higher. This
result suggests that the observations could be reproduced only if 
the back-reaction of accelerated CRs is extreme (i.e. $\gamma_{\rm eff}
\approx 1.1$) at all obliquity angles.

\begin{figure}[t!]
\resizebox{\hsize}{!}{\includegraphics[clip=true]{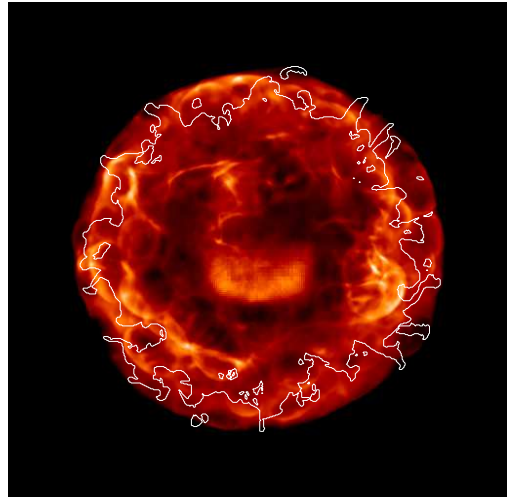}}
\caption{\footnotesize
3D rendering of mass density for a model with an exponential profile of
ejecta after 1000 yrs of evolution. The initial clumps have a size of
the order of 4\% of the initial remnant radius and a maximum density
contrast of 5. The white contour encloses the ejecta material.
}
\label{fig3}
\end{figure}

As a next step, we investigated the effects of ejecta clumping on the
evolution and morphology of the remnant, considering models without
back-reaction of accelerated CRs and accounting for ejecta clumping.
As an example, Fig.~\ref{fig3} shows the 3D rendering of mass density
for a model with an exponential profile of ejecta. The figure shows that:
1) the RT mixing reaches the forward shock front perturbing the remnant
outlines, 2) clumps and filamentary structures are evident within the
remnant, and 3) clumps of ejecta are close to or protruding beyond the
main blast wave as observed in SN\,1006 by \citet{Rakowski11}. In general,
we found that increasing the initial size of the clumps or increasing
their density contrast, both the perturbation of the remnant outlines
and the occurrence of ejecta protrusions increase. Also in this case,
we compared the azimuthal profiles of the ratio $R_{\rm bw}/R_{\rm cd}$
derived from the models with that observed in SN\,1006. We found that
the initial clumping of ejecta makes the azimuthal profiles $R_{\rm
bw}/R_{\rm cd}$ fairly uniform and lower than expected for models
without clumping. In particular, we found that: the larger the size of
initial clumps of ejecta, the lower the value of the ratio; the higher
the initial density contrast, the lower the value of the ratio.

As an example, in Fig.~\ref{fig4} we compare the azimuthal profiles
derived for models accounting for only one of the effects considered in
this paper (either back-reaction of accelerated CRs or ejecta clumping),
and for a model without both physical effects. Our analysis has shown
that the observed profile can be reproduced by models with a maximum
density contrast of ejecta ranging between 2.5 and 5, and an initial
size of ejecta clumps of the order of 4\% of the remnant radius (see,
for instance, blue line in Fig.~\ref{fig4}).

\begin{figure}[t!]
\resizebox{\hsize}{!}{\includegraphics[clip=true]{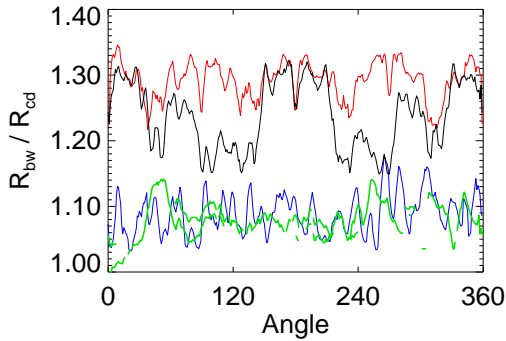}}
\caption{\footnotesize
Azimuthal profiles of the ratio between the forward shock and the contact
discontinuity radii for models without ejecta clumping and back-reaction
of accelerated CRs (M1, red line), with ejecta clumping and without
accelerated CRs (M2, blue), and with accelerated CRs and without ejecta
clumping (M3, black). All models assume an exponential profile of the
initial ejecta density; M2 assumes a minimum $\gamma_{\rm eff} = 1.1$;
M3 assumes a maximum density contrast 2.5 and an initial size of ejecta
4\% of the remnant radius. The green line marks the profile observed in
SN\,1006 (\citealt{Miceli09}).
}
\label{fig4}
\end{figure}

\section{Summary and conclusions}

We investigated the role of ejecta clumping and back-reaction of
accelerated CRs on the evolution of young SNRs.  To this end, we have
developed a 3D MHD model including consistently the back-reaction of
accelerated CRs and the ejecta clumping.  Particular attention has
been devoted to the spatial resolution that is necessary to describe
appropriately the structure of the ejecta; to this end, we have exploited
the AMR capabilities of the FLASH code. Our starting point was the
observational evidence that the azimuthal profile of the ratio between
the forward shock and the contact discontinuity radii is fairly uniform
and lower than expected for a non-modified shock.

We found that the back-reaction of accelerated CRs alone cannot reproduce
the observations unless the CRs energy losses are extreme (that is the
effective adiabatic index should be of the order of 1.1) and independent
on the obliquity angle. We also found that the clumping of ejecta is an
important factor to reproduce the observed values of the ratio and its
obliquity dependence. We conclude that the ejecta clumping is important
in the description of the evolution of young SNRs.

\begin{acknowledgements}
The software used in this work was in part developed by the DOE-supported
ASC / Alliance Center for Astrophysical Thermonuclear Flashes at
the University of Chicago. We acknowledge the CINECA Awards N.
HP10CYKB3A,2010 and HP10C7MTR0,2011 for the availability of high
performance computing resources and support. M.L.P. acknowledges
financial support from the Bonino-Pulejo Foundation.
\end{acknowledgements}

\bibliographystyle{aa}

\end{document}